\documentclass[12pt]{article}
\usepackage{a4}
\usepackage{amsmath}
\usepackage{amsfonts}
\usepackage{latexsym}
\usepackage{epsfig}

\begin{document}

\begin{flushright}
hep-th/0306232
\end{flushright}
\vskip 2 cm
\begin{center}
{\Large {\bf Softness of Brane-localized Supersymmetry Breaking
on Orbifolds}
}
\\[0pt]

\bigskip
\bigskip {\large
{\bf Ki-Young\ Choi ${}^{a,b,}$\footnote{
{{ {\ {\ {\ E-mail: ckysky@phya.snu.ac.kr}}}}}}},
{\bf Hyun Min\ Lee ${}^{b,}$\footnote{
{{ {\ {\ {\ E-mail: minlee@th.physik.uni-bonn.de}}}}}}}
\bigskip }\\[0pt]
\vspace{0.23cm}
{\it ${}^a$ School of Physics and Center for Theoretical Physics,} \\
{\it Seoul National University, Seoul 151-747, Korea.} \\
\vspace{0.23cm}
{\it ${}^b$ Physikalisches Institut der Universit\"at Bonn,} \\
{\it Nussallee 12, 53115 Bonn, Germany.}\\

\bigskip
\vspace{3.4cm} Abstract
\end{center}
We consider the brane-localized supersymmetry breaking
in 5D compactified on $S^1/Z_2$. In case of a bulk gaugino 
with arbitrary brane masses for its even and odd modes, we find the
mass spectrum and the wave functions of gaugino. 
We show that the gaugino masses at the distant brane are soft in the usual 
sense in the effective field theory with zero modes of bulk
gauge fields and they are also extremely soft in view of the
one-loop finite mass of a brane scalar in the KK regularization.
\vskip 6mm



\newpage

\section{Introduction}

\noindent
Orbifold compactification of extra dimensions is
necessary to get a chiral fermion and a lower supersymmetry
as zero modes from higher dimensions\cite{orbifolds}.
Moreover, in recent works on GUT orbifolds,
Scherk-Schwarz twists\cite{SS} have
been also used to break the GUT symmetry in higher dimensions 
into the SM gauge group
and break further the remaining supersymmetry after orbifolding.
It is noticeable that as far as the mass spectrum and the mode functions are
concerned, a Scherk-Schwarz(SS) breaking in orbifolds represented 
by a local symmetry in the Lagrangian is equivalent to
a Wilson-line breaking along extra dimensions\cite{gaugewl,gravitywl}. 
For instance, a SS twist for gauge symmetry breaking in orbifolds corresponds 
to a Wilson line of $\langle A_5 \rangle\neq 0$ of the 5D gauge field,
and a SS twist for supersymmetry breaking in orbifolds corresponds 
to a Wilson line of $\langle V^1_5+iV^2_5 \rangle\neq 0$ 
of the $SU(2)_R$ gauge fields 
in the 5D off-shell supergravity\cite{5doffshell}, 
which is the nonzero $F$ term of the radion multiplet\cite{pomarol}. 
There has been a lot of discussion on the softness 
of SS breaking of supersymmetry in 5D compactified
on the orbifold in view of the one-loop corrections for the zero mode 
of a bulk scalar\cite{quiros1,bhn,gn,delgado,pilo,hdkim}. 
It has been shown that the one-loop finiteness of SS breaking 
mainly comes from the so called KK regularization\cite{anton,quiros1,bhn,pilo}.

As an alternative to the Scherk-Schwarz breaking of supersymmetry, 
in this paper, we consider the brane-localized supersymmetry 
breaking\cite{yama,mirabelli,anomaly,gaugino,ah,evenmass,nilles2,nilles,quiros,ckl}. 
For simplicity, we consider the 5D SUSY $U(1)$ gauge theories on $S^1/Z_2$ 
where the brane-localized supersymmetry breaking is parametrized 
by general brane mass terms for gaugino.
When one introduces brane mass terms for gaugino, it is likely
to simply drop the mass term for the odd mode 
of gaugino\cite{pomarol,evenmass}.  However, in case
that the wave functions of odd modes have a discontinuity on the branes, the
odd mass term also contributes to the equations of motion so that it makes the
wave functions of even modes discontinuous 
on the branes\cite{nilles2,nilles,quiros}.
Then, the brane coupling
of the even modes are determined from the careful integration of
the brane action, but not from the equations of motion.

In this paper, with general brane mass terms for gaugino,
we find the mass spectrum and the wave functions of gaugino. 
While the mass spectrum is the same as the case with a specific 
Scherk-Schwarz parameter,
the wave functions of gaugino are modified due to the brane mass terms.
Therefore, we find that the generic brane mass terms are not
soft even in the usual sense in the effective field theory with zero modes
of gauge fields. 
We also show that for the same brane couplings of gauge boson and gaugino,
the one-loop finiteness of a brane scalar mass in our model 
is guaranteed in the KK regularization scheme. 
We find that this is the case with distant breaking
of supersymmetry\cite{mirabelli,anomaly,gaugino,ah}, 
i.e. brane matters at one brane and
only brane masses of gaugino at the other brane. 
The one-loop finiteness in our model is due to the distant  
supersymmetry breaking which is necessary for the 4D supersymmetric 
gauge coupling at the brane where matter fields are located. 

This paper is organized as follows. For comparison with our brane-localized 
supersymmetry breaking, we first give a brief review on the 
Scherk-Schwarz boundary condition in 5D compactified on $S^1/Z_2$.
In the section 3, we consider the general brane-localized supersymmetry breaking
in the gauge sector and show the wave functions and the mass spectrum 
of the bulk gaugino. Then, in the section 4, 
we present the one-loop KK gauge corrections 
to a massless scalar located at the brane and discuss its finiteness in the 
context of the distant supersymmetry breaking. 
In the last section, the conclusion is drawn.

\section{Scherk-Schwarz boundary conditions}

Let us first give a review on the Scherk-Schwarz breaking on orbifolds.
One can impose a general SS boundary condition on a bulk field $\Phi(x,y)$
living in $S^1$ with the radius $R$ as
\begin{eqnarray}
\Phi(x,y+2\pi R)=e^{2\pi i \omega}\Phi(x,y)
\end{eqnarray}
where $x,y$ denotes 4D and extra dimension coordinates respectively
and $\omega$ is the SS parameter.
Then, one gets a mode expansion of the bulk field as
\begin{eqnarray}
\Phi(x,y)=\frac{1}{\sqrt{2\pi R}}
\sum^{\infty}_{n=-\infty}e^{i(n+\omega)y/R}\Phi^{(n)}(x).
\end{eqnarray}
After the $Z_2$ orbifolding, which identifies $y$ with $-y$ in $S^1$,
the bulk field becomes even or odd under $Z_2$ as follows
\begin{eqnarray}
\Phi_+(x,y)&=&\frac{1}{2}(\Phi(x,y)+\Phi(x,-y)) \nonumber \\
&=&\frac{1}{\sqrt{2\pi R}}
\sum^{\infty}_{n=-\infty}\cos((n+\omega)y/R)\Phi^{(n)}(x), \label{emode}\\
\Phi_-(x,y)&=&\frac{1}{2i}(\Phi(x,y)-\Phi(x,-y)) \nonumber \\
&=&\frac{1}{\sqrt{2\pi R}}
\sum^{\infty}_{n=-\infty}\sin((n+\omega)y/R)\Phi^{(n)}(x)\label{omode}
\end{eqnarray}
with the mass spectrum
\begin{eqnarray}
M^2_n=\frac{(n+\omega)^2}{R^2}, \ \ \ n={\rm integer}.
\end{eqnarray}
Therefore, from eqs.~(\ref{emode}) and (\ref{omode}), 
we can rewrite the SS boundary conditions on $S^1/Z_2$ as
\begin{eqnarray}
\left(\begin{array}{l} \Phi_+ \\ \Phi_- \end{array}\right)(x,y+2\pi R)
=\left(\begin{array}{ll}
\cos(2\pi \omega) & -\sin(2\pi \omega) \\
\sin(2\pi \omega) & \cos(2\pi \omega)
\end{array}\right)
\left(\begin{array}{l}\Phi_+ \\ \Phi_- 
\end{array}\right)(x,y).
\end{eqnarray}

For instance, in 5D SUSY $U(1)$ gauge theories compactified on $S^1/Z_2$, 
the bulk gaugino is composed of two Weyl spinors $\lambda_1$ and $\lambda_2$, 
which are even and odd under $Z_2$, respectively. 
Then, performing a SS twist of the bulk gaugino 
and replacing the twisted gaugino by the untwisted bulk
gaugino(${\tilde\lambda}_1,{\tilde\lambda}_2$) as
\begin{eqnarray}
\left(\begin{array}{l} \lambda_1 \\ \lambda_2 \end{array}\right)(x,y)
=\left(\begin{array}{ll}
\cos(\omega y/R) & -\sin(\omega y/R) \\
\sin(\omega y/R) & \cos(\omega y/R)
\end{array}\right)
\left(\begin{array}{l}{\tilde \lambda}_1 \\ {\tilde\lambda}_2
\end{array}\right)(x,y),
\end{eqnarray}
one can show that the twisted bulk gaugino without mass terms
is equivalent to the untwisted bulk gaugino with constant bulk mass terms 
such as
\begin{eqnarray}
-\frac{1}{2}(\lambda_1\partial_y\lambda_2
-\lambda_2\partial_y\lambda_1)
=-\frac{1}{2}\frac{\omega}{R}
({\tilde \lambda}_1{\tilde \lambda}_1+{\tilde\lambda}_2{\tilde\lambda}_2).
\end{eqnarray}

\section{Brane-localized supersymmetry breaking}

Now we are in a position to consider the brane-localized supersymmetry 
breaking.
We consider a 5D SUSY $U(1)$ model compactified on $S^1/Z_2$ 
with the radius of $R$. After orbifolding,  
there appear two fixed points at $y=0$ and $y=\pi R$ 
where brane matters can be located. 
The 5D action for the bulk gaugino we are considering is
\begin{eqnarray}
S&=&\int d^4 x \int^{\pi R}_{-\pi R}dy
\bigg[\overline{\lambda_1}i{\bar \sigma}^\mu
\partial_\mu\lambda_1+\overline{\lambda_2}i{\bar \sigma}^\mu
\partial_\mu\lambda_2
-\frac{1}{2}(\lambda_1\partial_y\lambda_2
-\lambda_2\partial_y\lambda_1)+h.c. \nonumber \\
&-&\varepsilon_0(\lambda_1\lambda_1+\rho_0\lambda_2\lambda_2)\delta(y)
-\varepsilon_{\pi}(\lambda_1\lambda_1+\rho_{\pi}\lambda_2\lambda_2)
\delta(y-\pi R)+h.c.\bigg]\label{action}
\end{eqnarray}
where $\epsilon_{0,\pi}$ are the dimensionless parameters of brane mass terms
for gauginos and $\rho_{0,\pi}$ are the ratios between brane mass
parameters of even and odd modes of gaugino at each brane. 
The brane mass terms have been also considered 
only at one fixed point in $S^1/Z_2$ 
in the presence of the Scherk-Schwarz breaking\cite{quiros}.
In our case, 
we consider a more general situation 
where brane mass terms exist at both two fixed points in $S^1/Z_2$.

We have chosen two Weyl components of the bulk gaugino,
$\lambda_1$ and $\lambda_2$, to be even and odd
under $Z_2$ respectively as the following
\begin{eqnarray}
\lambda_1(-y)=\lambda_1(y), \ \ \ \lambda_2(-y)=-\lambda_2(y).
\end{eqnarray}
Then, when we make a KK reduction of the gaugino as
\begin{eqnarray}
\left ( \begin{array}{l}
\lambda_1(x,y) \\ \lambda_2(x,y)
\end{array} \right )
=\sum_n N_n \left ( \begin{array}{l}
u^{(n)}_1(y) \\ u^{(n)}_2(y)
\end{array} \right ) \lambda^{(n)}(x)
\end{eqnarray}
where $i{\bar\sigma}^\mu\partial_\mu\lambda^{(n)}=M_n\overline{\lambda^{(n)}}$
with the KK mass $M_n$ and $N_n$ is the normalization constant,
the equations of motion for the gaugino become
\begin{eqnarray}
\partial_y u^{(n)}_1+(M_n-2\rho_0\varepsilon_0\delta(y)
-2\rho_\pi\varepsilon_\pi\delta(y-\pi R))u^{(n)}_2&=&0,\label{eq1}\\
-\partial_y u^{(n)}_2+(M_n-2\varepsilon_0\delta(y)
-2\varepsilon_\pi\delta(y-\pi R))u^{(n)}_1&=&0.\label{eq2}
\end{eqnarray}
Now one can find it easy to solve the
equation for the ratio $t_n\equiv u^{(n)}_2/u^{(n)}_1$ as follows
\begin{eqnarray}
\frac{\partial t_n}{\partial y}=M_n(1+t^2_n)
-2\varepsilon_0(1+\rho_0 t^2_n)\delta(y)
-2\varepsilon_\pi (1+\rho_\pi t^2_n)\delta(y-\pi R).\label{eq3}
\end{eqnarray}
Thus, after integrating both sides of the above equation over an infinitesimal
interval around the branes, we obtain the following limiting values 
of $t_n$ on the boundaries
\begin{eqnarray}
\frac{1}{\sqrt{\rho_0}}{\rm arctan}(\sqrt{\rho_0}t_n)|_{y=0^+}
&=&-\varepsilon_0, \\
\frac{1}{\sqrt{\rho_\pi}}{\rm arctan}(\sqrt{\rho_\pi}t_n)|_{y=\pi R^-}
&=&\varepsilon_\pi.
\end{eqnarray}
Then, we get solutions for $t_n$ as
\begin{eqnarray}
t_n=\left\{\begin{array}{l}
\tan[M_n y-{\rm arctan}\,\alpha(\rho_0,\varepsilon_0\epsilon(y))], 
\ \ \ -\pi R<y<\pi R,\\
\tan[M_n (y-\pi R)-{\rm arctan}\,
\alpha(\rho_\pi,\varepsilon_\pi\epsilon(y-\pi R))], \ \ \ 0<y<2\pi R,
\end{array} \right.
\end{eqnarray}
where
\begin{eqnarray}
\alpha(\rho_0,\varepsilon_0\epsilon(y))
&=&\frac{1}{\sqrt{\rho_0}}\tan(\sqrt{\rho_0}\varepsilon_0\epsilon(y)),
\label{alpha1}\\
\alpha(\rho_0,\varepsilon_\pi\epsilon(y-\pi R))
&=&\frac{1}{\sqrt{\rho_\pi}}
\tan(\sqrt{\rho_\pi}\varepsilon_\pi\epsilon(y-\pi R)),\label{alpha2}
\end{eqnarray}
with $\epsilon(y)$ being the step function of periodicity
$2\pi R$ given by
\begin{eqnarray}
\epsilon(y)=\left\{ \begin{array}{l}
+1, \ \ \ 0<y<\pi R\\ 0, \ \ \ y=0 \\ -1, \ \ \ -\pi R<y<0.
\end{array}\right.
\end{eqnarray}
Here we note $\alpha(\rho_0,\varepsilon_0)={\rm tanh}(\sqrt{|\rho_0|}\varepsilon_0)/\sqrt{|\rho_0|}$ for $\rho_0<0$ and 
$\alpha(\rho_\pi,\varepsilon_\pi)={\rm tanh}(\sqrt{|\rho_\pi|}\varepsilon_\pi)/\sqrt{|\rho_\pi|}$ for $\rho_\pi<0$. 
We also find the mass spectrum of the gaugino as
\begin{eqnarray}
M_n=\frac{n}{R}+\frac{1}{\pi R}\bigg({\rm arctan}\,\alpha(\rho_0,\varepsilon_0)
+{\rm arctan}\,\alpha(\rho_\pi,\varepsilon_\pi)\bigg)\label{mass}
\end{eqnarray}
where $n$ is an integer. 
The mass spectrum with $\alpha(\rho_\pi,\varepsilon_\pi)=0$, 
i.e. $\varepsilon_\pi=0$, is the same as the result in Ref.~\cite{quiros}. 
Thus, we find that the mass spectrum of gaugino
is shifted by the amount given in terms of the brane mass parameters.
This is equivalent to the one from a Scherk-Schwarz breaking of parameter  
\begin{eqnarray}
\omega=\frac{1}{\pi}[{\rm arctan}\,\alpha(\rho_0,\varepsilon_0)
+{\rm arctan}\,\alpha(\rho_\pi,\varepsilon_\pi)]. \label{ssp}
\end{eqnarray}  
Particularly,
for $\alpha(\rho_0,\varepsilon_0)=-\alpha(\rho_\pi,\varepsilon_\pi)$, we have
the remaining supersymmetry restored. This would be the case with
two fine-tunings of $\varepsilon_0=-\varepsilon_\pi$ and $\rho_0=\rho_\pi$. 

For the strong supersymmetry breaking, $\varepsilon_0\gg 1$ and/or 
$\varepsilon_\pi\gg 1$, the mass spectrum depends on the sign of $\rho_0$ 
and $\rho_\pi$. For positive sign of odd-mode mass parameters, 
depending on the large even-mode mass parameters, 
the zero-mode gaugino mass oscillates between two values: 
$M_0\simeq \pm 1/R$ for $\rho_0>0$ and $\rho_\pi>0$ 
in the case with strong supersymmstry breaking on both branes while 
$M_0\simeq (\pm \frac{1}{2}+\frac{1}{\pi}{\rm arctan}\,\alpha(\rho_{\pi(0)},\varepsilon_{\pi(0)}))/R$ for $\rho_{0(\pi)}>0$
in the case with strong supersymmetry breaking on either brane.
On the other hand, for negative sign of odd-mode mass parameters, 
the leading mass spectrum
becomes independent of the large even-mode mass parameter 
but the still depends on $\rho_0$ and/or $\rho_\pi$: 
$M_0\simeq [{\rm arctan}(1/\sqrt{|\rho_0|})+{\rm arctan}(1/\sqrt{|\rho_\pi|})]/(\pi R)$ for $\rho_0<0$ and $\rho_\pi<0$
in the case with strong supersymmetry breaking on both branes
while $M_0\simeq [{\rm arctan}(1/\sqrt{|\rho_{0(\pi)}|})
+{\rm arctan}\,\alpha(\rho_{\pi(0)},\varepsilon_{\pi(0)})]/(\pi R)$ 
for $\rho_{0(\pi)}<0$ in the case with strong supersymmetry breaking 
on either brane.

Moreover, from the equations (\ref{eq1}) and (\ref{eq2}),
we get the eigen modes for the gaugino for $-\pi R<y<\pi R$ as follows
\begin{eqnarray}
\left ( \begin{array}{l}
u^{(n)}_1(y) \\ u^{(n)}_2(y)
\end{array} \right )= A(\rho_0,\varepsilon_0\epsilon(y))
\left ( \begin{array}{l}
\cos[M_n y-{\rm arctan}\,\alpha(\rho_0,\varepsilon_0\epsilon(y))] \\
\sin[M_n y-{\rm arctan}\,\alpha(\rho_0,\varepsilon_0\epsilon(y))]
\end{array} \right ) \label{sol}
\end{eqnarray}
where
\begin{eqnarray}
A(\rho_0,\varepsilon_0\epsilon(y))\equiv
\bigg(\frac{1+\alpha^2(\rho_0,\varepsilon_0\epsilon(y))}
{1+\rho_0\alpha^2(\rho_0,\varepsilon_0\epsilon(y))}\bigg)^{1/2}.
\end{eqnarray}
The prefactor $A(\rho_0,\varepsilon_0\epsilon(y))$ has been already found 
in Ref.~\cite{quiros}. 
However, for the analysis of brane couplings of gaugino, we need
to know the correct normalization constant which is obtained   
by inserting the equations of motion in the action as 
\begin{eqnarray}
N_n=\bigg(\int^{\pi R}_{-\pi R}dy[(u^{(n)}_1)^2+(u^{(n)}_2)^2]\bigg)^{-1/2}
=\frac{1}{\sqrt{2\pi R}}\frac{1}{A(\rho_0,\varepsilon_0)}.
\end{eqnarray}
Likewise, we get the eigen modes for the gaugino for $0<y<2\pi R$ as follows
\begin{eqnarray}
\left ( \begin{array}{l}
u^{(n)}_1(y) \\ u^{(n)}_2(y)
\end{array} \right )&=& (-1)^n 
A(\rho_\pi,\varepsilon_\pi\epsilon(y-\pi R))\times
\nonumber \\
&\times& \left ( \begin{array}{l}
\cos[M_n (y-\pi R)-{\rm arctan}\,
\alpha(\rho_\pi,\varepsilon_\pi\epsilon(y-\pi R))] \\
\sin[M_n (y-\pi R)-{\rm arctan}\,
\alpha(\rho_\pi,\varepsilon_\pi\epsilon(y-\pi R))]
\end{array} \right ) \label{sol2}
\end{eqnarray}
with the normalization constant 
\begin{eqnarray}
N_n=\bigg(\int^{2\pi R}_0 dy[(u^{(n)}_1)^2+(u^{(n)}_2)^2]\bigg)^{-1/2}
=\frac{1}{\sqrt{2\pi R}}\frac{1}{A(\rho_\pi,\varepsilon_\pi)}
\end{eqnarray}
where we inserted $(-1)^n$ in comparison with the previous solutions
for $0<y<\pi R$. 

Then, the values of even and odd mode functions of gaugino at the branes
are given by the definition
of $\epsilon(y)$ as $u^{(n)}_1(0)=1$, $u^{(n)}_1(\pi R)=(-1)^n$ 
and $u^{(n)}_2(0)=u^{(n)}_2(\pi R)=0$ in any case.
However, one should be careful in finding the {\it real} brane coupling
of gaugino with the integration of the product of a discontinous
mode function and a delta function.
The brane coupling of the $n$th($n$ is a nonnegative integer) 
KK mode of the bulk gauge 
boson\footnote{The loop correction coming from each massive KK mode with the brane coupling squared of $2g^2_4$ corresponds to those from two extra momentum states with the brane coupling squared of $g^2_4$. } is given as 
$\sqrt{2^{(1-\delta_{n,0})}}g_4$ at $y=0$ 
and $(-1)^n \sqrt{2^{(1-\delta_{n,0})}}g_4$ at $y=\pi R$ 
where $g_4=g_5/\sqrt{2\pi R}$.
On the other hand, the brane couplings of the $n$th($n$ is an integer) 
even mode of gaugino at $y=0$ and $y=\pi R$ are given
from the integrations of the brane action, respectively,
\begin{eqnarray}
g_0&\equiv&g_5\int dy \,\delta(y)N_n u^{(n)}_1(y) \nonumber \\
&=&g_4
A^{-1}_0\frac{\sin(\sqrt{\rho_0}\varepsilon_0)}{\sqrt{\rho_0}\varepsilon_0}
\label{coupling1}
\end{eqnarray}
and
\begin{eqnarray}
g_\pi&\equiv&g_5\int dy \,\delta(y-\pi R) N_n u^{(n)}_1(y)\nonumber \\
&=&g_4(-1)^n A^{-1}_\pi\frac{\sin(\sqrt{\rho_\pi}\varepsilon_\pi)}
{\sqrt{\rho_\pi}\varepsilon_\pi} \label{coupling2}
\end{eqnarray}
where 
$A_0\equiv A(\rho_0,\varepsilon_0)$ and 
$A_\pi\equiv A(\rho_\pi,\varepsilon_\pi) $. 
Of course, the brane couplings of the odd modes of
gaugino turn out to be zero after the integration of the brane action. 

For generic $\rho_{0,\pi}$ and $\varepsilon_{0,\pi}$, 
the brane coupling squared of the gaugino is different 
from that of the gauge boson. Henceforth let us use the word 
of {\it brane coupling} for {\it brane coupling squared} without confusion.
Irrespective of the mass spectrum of gaugino, 
the same brane coupling of gauge boson and gaugino is necessary
for no quadratic divergence, i.e. softly broken supersymmetry in usual sense, 
for a brane scalar which is located at either brane\cite{ckl}.
However, since our mass spectrum of gaugino
is given as that of a Scherk-Schwarz twist, the same brane coupling of
gauge boson and gaugino would give rise to one-loop finiteness, i.e. 
{\it extreme} softness of brane-localized 
supersymmetry breaking, which is the case with the distant supersymmetry
breaking as will be seen in the next section.

Particularly, for $\rho_0=\rho_\pi=0$, which is the usual assumption 
in the literature\cite{pomarol,evenmass}, 
the brane couplings at $y=0$ and $y=\pi R$ 
are proportional to $1/(1+\varepsilon^2_0)$ and
$1/(1+\varepsilon^2_\pi)$, respectively. 
In this case, the mass spectrum of gaugino is
$M_n=n/R+({\rm arctan}\,\varepsilon_0+{\rm arctan}\,\varepsilon_\pi)/(\pi R)$,
which is the same result as in \cite{ckl}. 
Then, imposing the additional condition 
$\varepsilon_0=0$ or $\varepsilon_\pi=0$ 
is necessary for the same coupling at either brane. 
On the other hand, for the equal masses of even and odd modes, 
i.e. $\rho_0=\rho_\pi=1$\cite{nilles},
the brane couplings at $y=0$ and $y=\pi R$ 
are proportional to $(\sin\varepsilon_0)^2/\varepsilon^2_0$ and 
$(\sin\varepsilon_\pi)^2/\varepsilon^2_\pi$, respectively. 
In this case, the mass spectrum
of gaugino is given by $M_n=n/R+(\varepsilon_0+\varepsilon_\pi)/(\pi R)$,
which is different from the case with vanishing odd mass terms.
For the same brane coupling of gauge boson and gaugino at either brane, 
we need the condition $\varepsilon_0=0$ or $\varepsilon_\pi=0$ again.
Thus, for general $\rho_0$ and $\rho_\pi$, which then contributes to the shape
of wave functions and the mass spectrum, 
we can show that with the local supersymmetry breaking at the distant brane, 
the brane couplings of gauge boson and gaugino are the same 
at the other brane.

\section{One-loop corrections at the brane}

As far as the gauge interaction with brane matters is concerned, 
the only difference between the brane-localized breaking and the Scherk-Schwarz
breaking comes from the brane scalar-gaugino-brane fermion vertices. 
After reducing the relevant brane interaction of gaugino, 
we get in the mass eigenstates
\begin{eqnarray}
{\cal L}_5&\supset& \int dy\,
g_5[-\sqrt{2}iq_0\phi_0^\dag \lambda_1\psi_0\delta(y)
-\sqrt{2}iq_\pi\phi_\pi^\dag \lambda_1\psi_\pi\delta(y-\pi R)+h.c.] 
\nonumber \\
&=&\sum^{\infty}_{n=-\infty}
[-g_0q_0\sqrt{2}i\phi_0^\dag \lambda^{(n)}\psi_0
-g_\pi q_\pi\sqrt{2}i\phi_\pi^\dag\lambda^{(n)}\psi_\pi+h.c.]
\end{eqnarray}
where $(\phi_0,\psi_0)$ and $(\phi_\pi,\psi_\pi)$ are brane matter multiplets
at $y=0$ and $y=\pi R$, respectively, and $g_{0,\pi}$ given 
by eqs.~(\ref{coupling1}) and (\ref{coupling2}) are brane couplings of gaugino
and $q_{0,\pi}$ denote $U(1)$ charges 
of brane matters. For comparison, in the case with a Scherk-Schwarz twist, 
$g_0=g_4$ at $y=0$ which was used to show the one-loop finiteness 
of the mass of a brane scalar at $y=0$ from the infinite sum 
of KK modes\cite{quiros1}. 

Thus, due to brane masses of gaugino, 
the one-loop correction to the mass of a massless scalar $\phi_0$
at $y=0$\cite{mirabelli,quiros1,ah,ckl} becomes nonzero as 
\begin{eqnarray}
-im^2_{\phi_0}&=&4g^2_4 q^2_0\sum^{\infty}_{n=-\infty}
\int\frac{d^4 p}{(2\pi)^4}
\bigg[\frac{1}{p^2-(n/R)^2}-\frac{r_0}{p^2-(n+\omega)^2/R^2}\bigg]
\nonumber \\
&=&i\frac{g^2_4 q^2_0}{2\pi^2 R^2}\sum^{\infty}_{n=-\infty}
\int^\infty_0 dx\, x^3
\bigg[-\frac{1}{x^2+n^2}+\frac{r_0}{x^2+(n+\omega)^2}\bigg]
\end{eqnarray} 
where $r_0\equiv g^2_0/g^2_4$, and $\omega$ given by eq.~({\ref{ssp}}) 
corresponds to a sort of SS parameter
and in the second line, 
we changed to the variable $x=p_E R$ with the Euclidean momentum $p_E$. 
Likewise, the one-loop correction to the mass of a massless scalar
$\phi_\pi$ at $y=\pi R$ is given by $m^2_{\phi_0}$ 
with $(g_0,q_0,r_0)\rightarrow (g_\pi,q_\pi,r_\pi=g^2_\pi/g^2_4)$. 
Then, with the $\Lambda$ cutoff regularization for the 4D loop integral 
at each KK level and the cutoff of the number of KK modes 
$N=[\Lambda R]$\cite{ckl}, we get the one-loop scalar mass as 
\begin{eqnarray}
m^2_{\phi_0}&=&\frac{g^2_4 q^2_0}{4\pi^2R^2}\sum^{N}_{n=-N}
\bigg[(1-r_0)(\Lambda R)^2-n^2\ln\frac{(\Lambda R)^2+n^2}{n^2}\nonumber \\
&+&r_0 (n+\omega)^2\ln\frac{(\Lambda R)^2+(n+\omega)^2}{(n+\omega)^2}\bigg].
\end{eqnarray}
Thus, for $r_0\neq 1$, the one-loop scalar mass at $y=0$ has 
a quadratic divergence as well as a log divergence at each KK level. 
In fact, $r_0\neq 1$ is not the supersymmetric gauge coupling
in the 4D effective field theory with softly broken supersymmetry.
For the small brane mass parameters, $\varepsilon_0\ll 1$, 
we get $r_0\simeq 1+(\frac{2}{3}\rho_0-1)\varepsilon^2_0+{\cal O}(\varepsilon^4_0)$ from eq.~(\ref{coupling1}), which gives rise to
the reduction of the sum of quadratic divergences  
with the cutoff of the number of KK modes\cite{ckl}. 

Now let us take a different regularization scheme for the loop divergence. 
When we can rewrite the one-loop scalar mass at $y=0$ as
\begin{eqnarray}
m^2_{\phi_0}=\frac{g^2_4 q^2_0}{2\pi^2 R^2}(C(0)-r_0 C(\omega))
\end{eqnarray}
where
\begin{eqnarray}
C(\omega)=\sum^{\infty}_{n=-\infty}\int^\infty_0 dx\, 
\frac{x^3}{x^2+(n+\omega)^2},
\end{eqnarray}
and change the infinite sum of KK modes in $C(\omega)$ 
into the contour integral\cite{mirabelli,quiros1}, we get  
\begin{eqnarray}
C(0)-r_0 C(\omega)&=&\frac{\pi}{2}\int^\infty_0 dx\, x^2
\bigg[{\rm coth}(\pi x)-r_0{\rm coth}(\pi(x+i\omega))
+h.c.\bigg] \nonumber \\
&=&\pi(1-r_0)\int^\infty_0 dx\, x^2 \nonumber \\
&+&\frac{1}{4\pi^2}\bigg[2\zeta(3)-
r_0({\rm Li}_3(e^{-2i\pi\omega})+{\rm Li}_3(e^{2i\pi\omega}))\bigg]
\end{eqnarray}
where $\zeta(3)$ is the Riemann's zeta function and ${\rm Li}_3(x)$ is 
the trilogarithm as 
\begin{eqnarray}
{\rm Li}_3(x)=\sum^{\infty}_{k=1}\frac{x^k}{k^3}.
\end{eqnarray}
Therefore, for $r_0\neq 1$, there would still appear a cubic divergent one-loop 
mass, which corresponds to the sum of quadratic divergences coming from KK 
modes. However, there is no other divergence in this regularization.  
For no cubic divergence in this regularization, 
we must take $r_0=1$, i.e. $\varepsilon_0=0$, 
for which the SS parameter is given by
$\omega={\rm arctan}(\alpha(\rho_\pi,\varepsilon_\pi))/\pi$. 
This is the case with gaugino mediation of supersymmetry
breaking at the distant brane\cite{gaugino}. 
In this case, the one-loop radiative mass squared 
for a brane scalar is positive and finite, 
which means that the brane-localized supersymmetry breaking is 
extremely soft in the so called KK regularization scheme. This infinite sum
of KK modes was advocated from the mixed position-momentum
propagator of the bulk field\cite{ah}. 
Likewise, a massless scalar $\phi_\pi$ at $y=\pi R$ also gets a similar finite
one-loop mass for $g_\pi=g_4$, i.e. $\varepsilon_\pi=0$, for which 
the corresponding SS parameter is given by 
$\omega={\rm arctan}(\alpha(\rho_0,\varepsilon_0))/\pi$.

This result also sheds light on the aspect of supersymmetric 
flavor problem.
In the presence of distant supersymmetry breaking in the gauge sector, 
we can generalize the result to the case with a bulk non-abelian group. 
Thus, we find that radiative 
soft masses of brane scalars are to be finite and flavor diagonal as
\begin{eqnarray}
(m^2_{\phi_0})^i_j&=& \delta^i_j\frac{g^2_4 C_2(\phi)}{4\pi^4 R^2}
\sum^\infty_{k=1}\frac{(1-\cos(2\pi k\omega))}{k^3} \nonumber \\
&\simeq &\delta^i_j\frac{g^2_4 C_2(\phi)}{\pi^2 }
\bigg(\frac{\omega}{R}\bigg)^2
\bigg[\frac{3}{4}-\frac{1}{2}\ln(2\pi\omega)\bigg]
\end{eqnarray}
where we picked up the leading term in powers of $\omega^2$ and 
$C_2(\phi)$ is the quadratic Casimir of the $\phi$-representation 
under the gauge group.

\section{Conclusion}

To conclude, we considered the brane-localized supersymmetry breaking
on $S^1/Z_2$ by introducing brane mass terms for the bulk gaugino.
We have found that the brane
mass terms for the odd mode of gaugino play a role in modifying the mass
spectrum of gaugino and determining the brane coupling of the even mode
of gaugino. We showed that in the presence of brane gaugino mass terms,
the mass spectrum of gaugino is shifted by the amount given by brane mass 
parameters.
For the local supersymmetry breaking at the distant brane, 
we found that KK gauge corrections to the self-energy of a brane scalar is
soft and flavor diagonal at one-loop order.

\section*{Acknowledgments}
We would like to thank Hyung Do Kim and Hans Peter Nilles for valuable comments 
on the manuscript. We also appreciate helpful comments from Marek Olechowski
at the early stage of our work.
KYC is supported in part by the BK21 program
of Ministry of Education, KOSEF Sundo
Grant, and Korea Research Foundation Grant No. KRF-PBRG-2002-070-C00022.
This work(HML) is supported by the
European Community's Human Potential Programme under contracts
HPRN-CT-2000-00131 Quantum Spacetime, HPRN-CT-2000-00148 Physics Across the
Present Energy Frontier and HPRN-CT-2000-00152 Supersymmetry and the Early
Universe. HML was supported by priority grant 1096 of the Deutsche
Forschungsgemeinschaft.


\begin{thebibliography}{99}
\def\apj#1#2#3{Astrophys.\ J.\ {\bf #1} (#2) #3}
\def\ijmp#1#2#3{Int.\ J.\ Mod.\ Phys.\ {\bf #1} (#2) #3}
\def\mpl#1#2#3{Mod.\ Phys.\ Lett.\ {\bf #1} (#2) #3}
\def\nat#1#2#3{Nature\ {\bf #1} (#2) #3}
\def\npb#1#2#3{Nucl.\ Phys.\ {\bf B #1} (#2) #3}
\def\plb#1#2#3{Phys.\ Lett.\ {\bf B #1} (#2) #3}
\def\prd#1#2#3{Phys.\ Rev.\ {\bf D #1} (#2) #3}
\def\prl#1#2#3{Phys.\ Rev.\ Lett.\ {\bf #1} (#2) #3}
\def\prt#1#2#3{Phys.\ Rep.\ {\bf #1} (#2) #3}
\def\sjnp#1#2#3{Sov.\ J.\ Nucl.\ Phys.\ {\bf #1} (#2) #3}
\def\zp#1#2#3{Z.\ Phys.\ {\bf C #1} (#2) #3}
\def\jhep#1#2#3{JHEP \ {\bf #1} (#2) #3}
\def\epjc#1#2#3{Eur.\ Phys.\ J. \ {\bf C #1} (#2) #3}



\bibitem{orbifolds} L. J. Dixon, J. A. Harvey, C. Vafa and E. Witten,
\npb{261}{1985}{678}; \npb{274}{1986}{285}.


\bibitem{SS} J. Scherk and J. H. Schwarz, \plb{82}{1979}{60};
\npb{153}{1979}{61}.

\bibitem{gaugewl} Y. Hosotani, \plb{126}{1983}{309};
Annals Phys. {\bf 190} (1989) 233; 
L. J. Hall, H. Murayama and Y. Nomura, \npb{645}{2002}{85}.

\bibitem{gravitywl} G. v. Gersdorff and M. Quiros, \prd{65}{2002}{064016}.


\bibitem{5doffshell} M. Zucker, \npb{570}{2000}{267}; \jhep{08}{2000}{016};
\prd{64}{2001}{024024}.


\bibitem{pomarol} D. Marti and A. Pomarol, \prd{64}{2001}{105025}.


\bibitem{quiros1} I. Antoniadis, S. Dimopoulos, A. Pomarol and M. Quiros,
\npb{544}{1999}{503} [hep-ph/9810410]; 
A. Delgado, A. Pomarol and M. Quiros, \prd{60}{1999}{095008} [hep-ph/9812489].

\bibitem{bhn}
R. Barbieri, L. J. Hall and Y. Nomura, \npb{624}{2002}{63} [hep-ph/0011311].

\bibitem{gn} D. M. Ghilencea and H. P. Nilles, \plb{507}{2001}{327} 
[hep-ph/0103151];
D. M. Ghilencea, H. P. Nilles and S. Stieberger, New J. Phys. {\bf 4} (2002) 15
[hep-th/0108183];
D. M. Ghilencea and H. P. Nilles, 
J. Phys. G {\bf 28} (2002) 2475 [hep-ph/0204261];
T. Kobayashi and H. Terao, Prog. Theor. Phys. {\bf 107} (2002) 785 
[hep-ph/0108072]. 

\bibitem{delgado} A. Delgado, G. v. Gersdorff, P. John and M. Quiros,
[hep-ph/0104112].

\bibitem{pilo}
R. Contino and L. Pilo, \plb{523}{2001}{347} [hep-ph/0104130].


\bibitem{hdkim}
H. D. Kim, \prd{65}{2002}{105021} [hep-th/0109101].


\bibitem{anton} I. Antoniadis, \plb{246}{1990}{317}.


\bibitem{yama} H. P. Nilles, M. Olechowski and M. Yamaguchi, 
\plb{415}{1997}{24} [hep-th/9707143];
\npb{530}{1998}{43} [hep-th/9801030].


\bibitem{mirabelli} E. A. Mirabelli and M. E. Peskin, \prd{58}{1998}{065002}.


\bibitem{anomaly} L. Randall and R. Sundrum, \npb{557}{1999}{79}; 
C. F. Giudice and R. Rattazzi, \jhep{9812}{1998}{027}. 


\bibitem{gaugino} D. E. Kaplan, G. D. Kribs and M. Schmaltz, 
\prd{62}{2000}{035010} [hep-ph/9911293]; 
Z. Chacko, M. A. Luty, A. E. Nelson and E. Ponton,
\jhep{0001}{2000}{003} [hep-ph/9911323]; See also for the warped extra 
dimension case, Y. Nomura and D. R. Smith, hep-ph/0305214.



\bibitem{ah} N. Arkani-Hamed, L. Hall, Y. Nomura, D. Smith and N. Weiner,
[hep-ph/0102090].


\bibitem{evenmass} J. A. Bagger, F. Feruglio and F. Zwirner,
\prl{88}{2002}{101601} [hep-th/0107128];
\jhep{0202}{2002}{010} [hep-th/0108010].


\bibitem{nilles2} K. A. Meissner, H. P. Nilles and M. Olechowski,
\npb{561}{1999}{30}.


\bibitem{nilles} K. A. Meissner, H. P. Nilles and M. Olechowski,
Acta Phys. Polon. {\bf B33} (2002) 2435 [hep-th/0205166].


\bibitem{quiros} A. Delgado, G. v. Gersdorff and M. Quiros,
\jhep{0212}{2002}{002} [hep-th/0210181].


\bibitem{ckl} K. -Y. Choi, J. E. Kim and H. M. Lee, hep-ph/0303213.












\end{thebibliography}
\end{document}